\documentclass{PoS}

\usepackage{graphicx,amsmath,amssymb,fontenc,times,mathptmx}

\usepackage{tikz}
\usetikzlibrary{shadows}
\usetikzlibrary{arrows,shapes,positioning}
\usetikzlibrary{decorations.markings}
\usetikzlibrary{decorations.pathmorphing}
\usetikzlibrary{decorations.pathreplacing}
\tikzstyle arrowstyle=[scale=1]
\tikzstyle directed=[postaction={decorate,decoration={markings, mark=at position .65 with {\arrow[arrowstyle]{stealth}}}}]
\tikzstyle end directed=[postaction={decorate,decoration={markings, mark=at position 1 with {\arrow[arrowstyle]{stealth}}}}]
\tikzstyle reverse directed=[postaction={decorate,decoration={markings, mark=at position .65 with {\arrowreversed[arrowstyle]{stealth};}}}]
\tikzstyle{ann} = [fill=white,font=\footnotesize,inner sep=1pt]

\title{First results with twisted mass fermions towards the computation of parton distribution functions on the lattice}

\ShortTitle{Hadron structure}

\author{Constantia Alexandrou$^{(a,b,c)}$,  Krzysztof Cichy$^{(c,d)}$, Vincent Drach$^{(e)}$, Elena Garcia-Ramos$^{(c,f)}$, Kyriakos Hadjiyiannakou$^{(a)}$, Karl Jansen$^{(c)}$, Fernanda Steffens$^{(c)}$, \speaker{Christian Wiese}$\;^{(c)}$\\
\\
        $^{(a)}\;$Department of Physics, University of Cyprus, P.O. Box 20537, 1678 Nicosia, Cyprus\\
        $^{(b)}\;$Computation-based Science and Technology Research Center, Cyprus Institute, 20 Kavafi Str., Nicosia 2121, Cyprus\\
        $^{(c)}\;$John von Neumann Institute for Computing (NIC), DESY, Platanenallee 6, D-15738 Zeuthen, Germany\\
	$^{(d)}\;$Faculty of Physics, Adam Mickiewicz University, Umultowska 85, 61-614 Pozna\'{n}, Poland \\
        $^{(e)}\;$CP$^3$-Origins and the Danish Institute for Advanced Study DIAS, University of Southern Denmark, Campusvej 55, DK-5230 Odense M, Denmark\\
        $^{(f)}\;$Humboldt-Universit{\"a}t zu Berlin, Institut f{\"u}r Physik, Newtonstra{\ss}e 15, D-12489 Berlin, Germany\\
	\\
        E-mail: \email{christian.wiese@desy.de}}

\abstract{We report on our exploratory study for the evaluation of the parton distribution functions
from lattice QCD, based on a new method proposed in Ref.~arXiv:1305.1539.  
Using the example of the nucleon, we compare two different methods to compute the matrix elements needed,
and investigate the application of gauge link smearing. We also present first results from a large
production ensemble and discuss the future challenges related to this method.}

\FullConference{The 32nd International Symposium on Lattice Field Theory,\\
		23-28 June, 2014\\
		Columbia University New York, NY}

\begin{document}

\section{Introduction}

An essential part in understanding QCD is to study the structure
of hadrons, e.g. the nucleon.
Therefore, in the last decades a lot of effort was spent into
obtaining the parton distribution function (PDF) of the nucleon.
In particular, $q(x)$ is the PDF that gives the probability 
of finding a quark $q$ carrying
a momentum fraction $x$ of the parent hadron.

On the experimental side, deep inelastic scattering is an important
tool to access the structure of the nucleon. From the scattering
cross section of these experiments structure functions of the
nucleon can be extracted. Perturbative QCD can then be used to relate
the experimental results obtained from different scales. However,
perturbation theory does not give us any information about the
quark and gluon distributions themselves. For that one needs
to assume input distributions, which are then fitted to the
available data set. Different groups make different assumptions
for the form of these input distribution and which 
data sets to include in the fit. This yields the PDFs albeit with some model dependence.

Therefore, a method to compute the quark distribution from first 
principles is highly desirable. On the lattice, however, it was
up to now only possible to compute low moments of the PDFs (cf. e.g. \cite{Alexandrou:2014yha}).

Recently, a method to compute these distributions on a Euclidean 
lattice has been proposed in \cite{Ji:2013dva}, and has first been
applied in \cite{Lin:2014zya}. We present an exploratory study of 
this method using a setup of maximally twisted mass fermions
with a focus on different methods to calculate the PDFs and
systematic errors appearing in the computations.

\section{Euclidean formulation of the light cone operator}

The quark distribution inside a nucleon is usually defined
via matrix elements of the light cone operator,
\begin{align}
\label{EQN_pdf}
q(x) = \frac{1}{2\pi}\int \text{d}\xi^-\text{e}^{-ixp^+\xi^-}\langle N \vert \overline{\psi}(\xi^-)\Gamma \mathcal L(\xi^-,0)\psi(0)\vert N\rangle,
\end{align}
where $\xi^- = \frac{\xi^0 - \xi^3}{\sqrt{2}}$ and $\mathcal L(\xi^-,0)$ is the Wilson line from 0 to $\xi^-$. 
Because deep inelastic scattering is light cone dominated
this expression has to be evaluated at $\xi^2 \sim 0$. 
On a Euclidean lattice we would thus have to select
$\xi^2=\vec{x}^2+t^2 \sim 0$, which is very hard to 
calculate due to a non-zero lattice spacing.

A new idea is to compute a quasi distribution $\tilde{q}$ 
which is purely spatial and uses nucleons with finite momentum;
hence, it can be computed on the lattice. The quasi distribution
is given by
\begin{align}
\label{EQN_quasi}
\tilde{q}(x,\mu^2,p^z) = \frac{1}{2\pi} \int d\Delta z\;e^{-ixp^z \Delta z}\langle N(p^z) \vert \overline{\psi}(\Delta z)\gamma^z \mathcal L(\Delta z,0)\psi(0)\vert N(p^z) \rangle_{\mu^2}.
\end{align}
Here $\Delta z$ is a distance in any spatial direction 
$z$ and $p^z$ a momentum boost in this direction.
Notice that when $p^z \rightarrow \infty$ we recover the usual 
definition of the quark distributions in the infinite momentum
frame, which are equivalent to the light cone distributions of
Eq.~(\ref{EQN_pdf}). However, on the lattice one is limited to finite values
of $p^z$. Thus, the proposal is to calculate $\tilde{q}$ at finite 
$p^z$ and relate it to the usual distributions via perturbation 
theory \cite{Xiong:2013bka},
\begin{align}
\label{EQN_matching}
\tilde{q}(x,\mu^2,p^z) = \int \frac{dy}{\vert y \vert}\; Z\left(\frac{x}{y},\frac{\Lambda_{QCD}}{p^z},\frac{\mu}{p^z}\right )q(y,\mu^2) + \mathcal O \left(\left(\frac{m_N}{p^z}\right)^2\right).
\end{align}

\section{Implementation and algorithmic tests}

This work presents the first steps of the PDF calculation, 
which is the computation of the matrix elements appearing in Eq.~(\ref{EQN_quasi}),
and some of the related algorithmic and conceptual challenges.

The first task is to compute the bare matrix element needed for the quasi distribution,
\begin{align}
h(p^z,\Delta z) =\left \langle N(p^z) \vert \overline{\psi}(\Delta z)\gamma^z \mathcal L(\Delta z,0)\psi(0)\vert N(p^z) \right\rangle.
\end{align}
In order to compute these matrix elements a
nucleon-nucleon three-point function is required:
\begin{align}
C^{\text{3pt}}(t,\tau,0) = \langle\Gamma_{\alpha\beta}N_{\alpha}(\vec{p}^z,t) \mathcal O(\tau) \overline{N}_{\beta}(\vec{p}^z,0)\rangle, 
\end{align} 
where $\Gamma_{\alpha\beta}$ is a suitable parity projector. 
Here we will use the parity plus projector $\Gamma = \frac{1+\gamma_4}{2}$.

A nucleon field boosted with a three momentum can be 
defined via a Fourier transformation of quark fields in position space
\begin{align}
N_{\alpha}(\vec{p}^z,t) = \sum_{\vec{x}}\text{e}^{i \vec{p}^z \vec{x}}\epsilon^{abc}u_\alpha^a(x)\left( {d^b}^T(x)\mathcal C \gamma_5 u^c(x)\right).
\end{align}

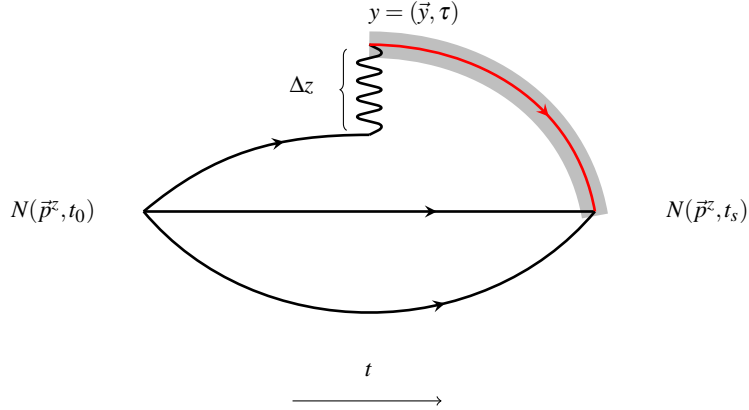
\begin{figure}
        \centering
	\begin{tikzpicture}[scale = 0.6]
\draw[gray!50, line width=10pt]  (5.0,3.7) to[out=0,in=100] (10,-0.1);
\draw[black, directed, line width=1pt] (0,0) -- (10,0);
\draw[black, line width=1pt]  (5,1.7) {[rounded corners] -- (5.4,1.9) -- (4.6,2.1) -- (5.4,2.3) -- (4.6, 2.5) -- (5.4, 2.7) -- (4.6,2.9) -- (5.4, 3.1) -- (4.6,3.3) --  (5.4,3.5) -- (5.0,3.7)};
\draw[red, directed,line width=1pt]  (5.0,3.7) to[out=0,in=100] (10,0);
\draw[black,directed, line width=1pt] (0,0) to[out=40,in=180]   (5,1.7);
\draw[black, directed,line width=1pt] (0,0) to[out=-50,in=-130] (10,0);
\node[ann] at (-2,0) {$N(\vec{p}^z,t_0)$};
\node[ann] at (12.5,0) {$N(\vec{p}^z,t_s)$};
\node[ann] at (6,4.4) {$y=(\vec{y},\tau)$};
\draw[arrows=->] (3.3,-4.2) -- (6.6,-4.2);
\draw[decorate, decoration={brace}] (4.5,1.8) -- (4.5,3.6);
\node[ann] at (3.5,2.8) {$\Delta z$};
\node[ann] at (5,-3.5) {$t$};
\end{tikzpicture}
	\caption{\label{FIG_DIA} Schematic picture of a possible Wick contraction of the quark fields in the three-point function.}
\end{figure}

In order to obtain a vanishing momentum transfer at the 
operator ($Q^2=0$) we need to write it as follows:
\begin{align}
\mathcal O(\Delta z, \tau, Q^2=0) = \sum_{\vec{y}}\overline{\psi}(y+\Delta z)\gamma^z \mathcal L(y+\Delta z,y)\psi(y),
\end{align}
with $y=(\vec{y},\tau)$. After Wick contraction of the quark
fields, the three-point function can be expressed in terms
of quark propagators. Fig.~\ref{FIG_DIA} shows a possible
contraction.

Finally we can extract the matrix element via a ratio of a three- and 
a two-point function $C^{\text{2pt}}(t,0;\vec{p}^z)=\Gamma_{\alpha\beta}N_{\alpha}(\vec{p}^z,t) \overline{N}_{\beta}(\vec{p}^z,0)$
\begin{align}
\frac{C^{\text{3pt}}(t,\tau,0;\vec{p}^z)}{C^{\text{2pt}}(t,0;\vec{p}^z)}\stackrel{0\ll \tau\ll t}{=}\frac{-ip^z}{E}h(p^z,\Delta z),
\end{align}
where $E=\sqrt{(p^z)^2+m_N^2}$ is the total energy of the
nucleon. As the flavor structure  of the
operator we will use $u-d$
in order to have an isovector operator and avoid disconnected diagrams.

An important conceptual problem one encounters when trying to compute the matrix element
is the calculation of the propagator connecting the sink with the operator insertion (marked red in Fig.~\ref{FIG_DIA}).
Due to momentum projection there is a spatial sum on both ends of the propagator.
For this purpose an all-to-all propagator is needed. However, the computation 
using point sources would take $V=L^3\times T$ sets of inversion, which
is too much computational effort.

Here we test two different approaches to compute the all-to-all propagator:
the sequential method, where we use standard methods to compute 
the two quark propagators that do not have the operator insertion and contract the result
to a sequential source. This can again be inverted and then contracted
with the operator to obtain the final result. However, in order to use this method 
the sink has to be fixed, which requires a new inversion
for each momentum. Thus, six sets of inversions have to be computed for each 
flavor, accounting for positive and negative momenta in all three spatial directions.
For higher momenta this would have to be repeated.

The second method is the stochastic method, where we use sources that contain 
$Z^4$ noise on one single time-slice (cf. e.g. \cite{Alexandrou:2013xon}). In addition we use diluted sources, which 
means that the noise is only on one single spin and color slice.
After inverting these sources, the obtained sink
can be used together with the source to estimate the all-to-all propagator.
As a drawback we add stochastic noise to the result, however, 
for this method one set of inversions for each flavor is sufficient to account for all momenta.

\begin{figure}
	\centering
	\includegraphics[scale=1]{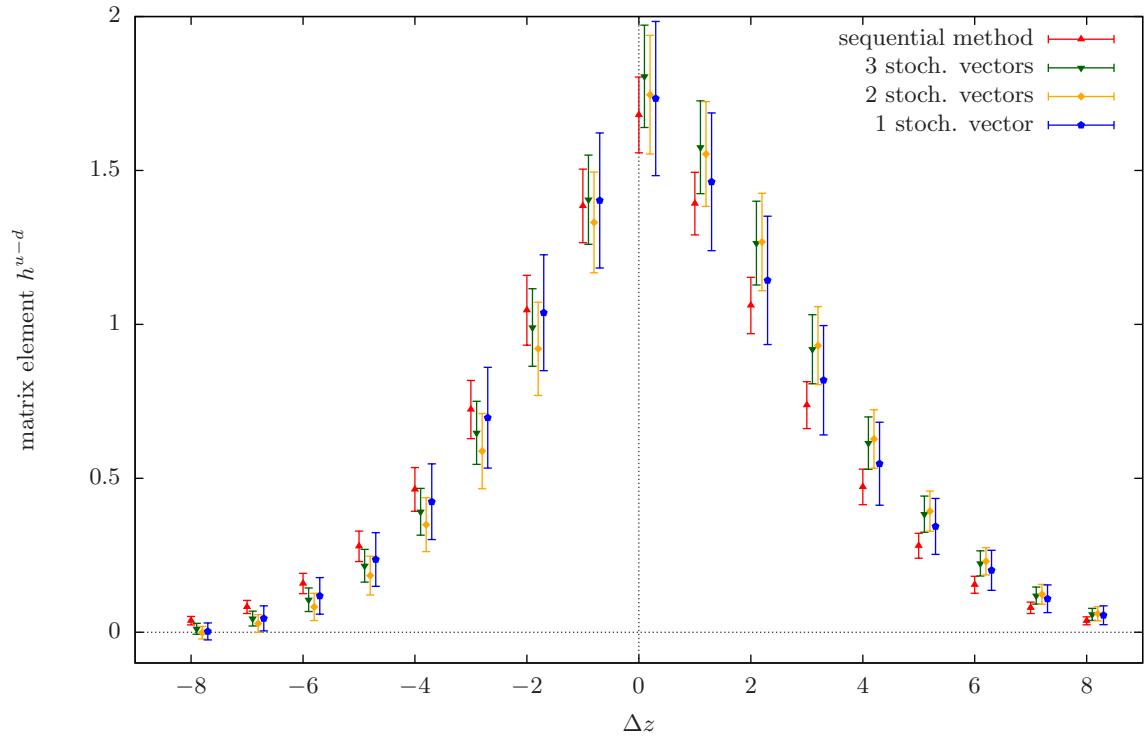}
	\caption{\label{FIG_SEQ_STOCH}Comparison of the sequential method with the stochastic method (with several numbers of noise vectors). The computational cost of the sequential method equals six noise vectors of the stochastic method. We show the real part of the matrix element.}
\end{figure}

For first initial tests we use a $16^3 \times 32$ twisted mass ensemble with $\beta=3.9$ \cite{Cichy:2012vg}
generated by the tmLQCD software package \cite{Jansen:2009xp}.
We choose a twisted mass parameter of $\mu = 0.004$, which corresponds to 
a pion mass of $m_{PS} \approx 340$\;MeV.
For the matrix element the source-sink separation is $6a$ and we only compute
momentum $p^z=1\frac{2\pi}{L}$. We show only the real part of the 
matrix elements. The imaginary part has a non-vanishing
contribution, however, it is relatively small and can at the 
moment be neglected. Fig.~\ref{FIG_SEQ_STOCH} shows the comparison between the
sequential and the stochastic method. For the stochastic method
we show a different number of noise vectors. Note that six noise
vectors would compare to the computational cost of the sequential
method. 

Both methods yield results that are compatible within errors. One cannot
directly compare the errors of both methods. However, taking into account
the different costs of inversions, both methods are approximately equal.

\begin{figure}
	\centering
	\includegraphics[scale=1]{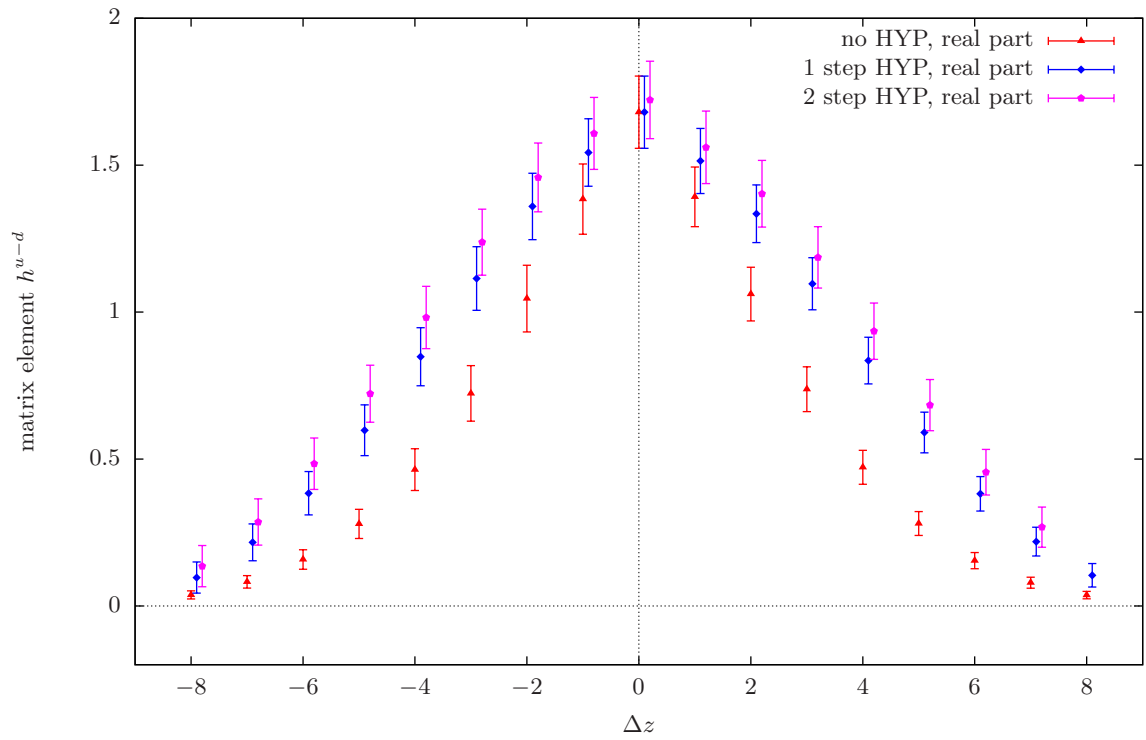}
	\caption{\label{FIG_HYP} Different number of gauge link smearing steps for the real part of the matrix element.}
\end{figure}

In \cite{Lin:2014zya} the authors applied HYP smearing \cite{Hasenfratz:2001hp} to the gauge
links. This is known to bring the necessary
renormalization factors closer to the corresponding tree level value.
To study the influence of HYP smearing we apply one and two
steps of smearing and show the results in Fig.~\ref{FIG_HYP}.

The results indicate that gauge link smearing increases the
value of the matrix elements, especially for larger $\Delta z$,
however, it does not decrease the noise of the result, as it was
observed e.g. for the gluon moment \cite{Alexandrou:2013tfa}.
Thus, for the following computations we will not use gauge link smearing and
try to compute the renormalization factor for each $\Delta z$  separately.

\section{Results from $N_f=2+1+1$ ensemble}

After these initial tests we continued with a larger 
ETMC production ensemble \cite{Baron:2010bv}. We decided to use the stochastic
method because, although both methods yield equal results,
the stochastic method is more flexible concerning the study
of larger momenta.
The matrix elements are computed on a $32^3 \times 64$ lattice
with $N_f=2+1+1$ flavors of maximally twisted mass fermions.
This ensemble has $\beta=1.95$, which corresponds to a lattice spacing of $a\approx 0.078$\;fm
and the twisted mass parameter $\mu = 0.0055$, which is a pion mass of
$m_{PS} \approx 373$\;MeV. All the results presented are computed
with a source-sink separation of $10a$.
\begin{figure}
	\centering
	\includegraphics[scale=1]{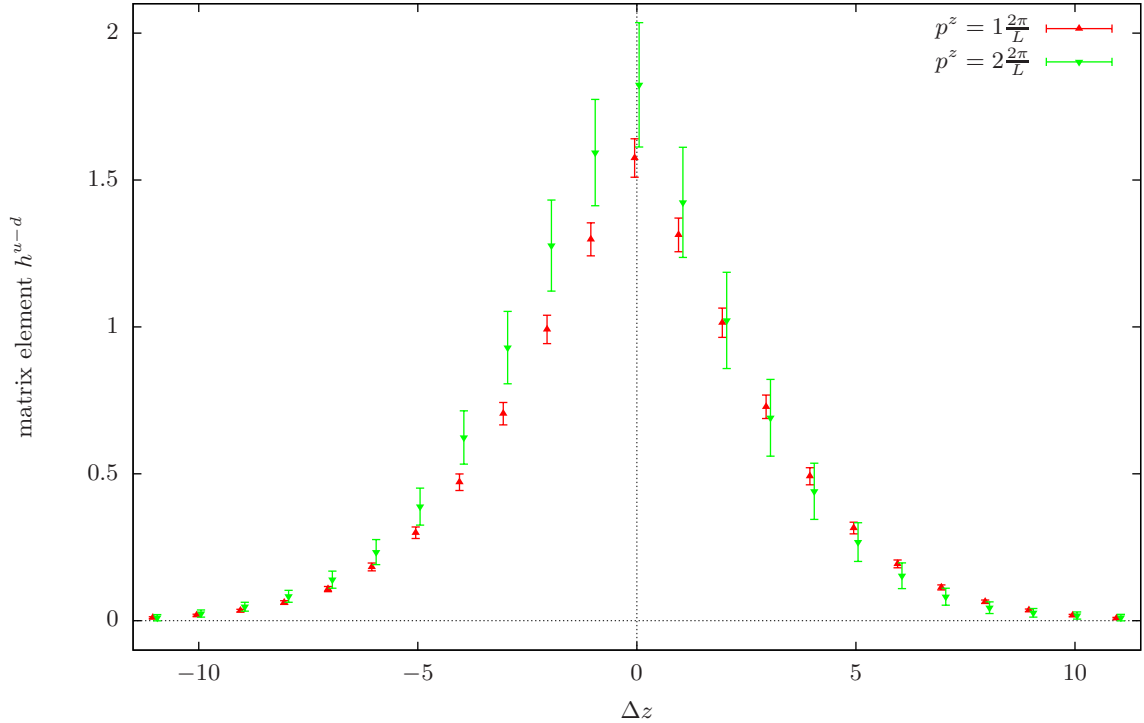}
	\caption{\label{FIG_B55} Real part of the matrix element for the first two momenta with 1000 measurements.}
\end{figure}
With our current statistics of $N_{conf}=1000$ we are able to extract the matrix
element for the first two momenta. We display the result in Fig.~\ref{FIG_B55}.
Note that the value for $\Delta z = 0$, which can be identified with 
the local vector current at $Q^2=0$, has to be renormalized with $Z_V$, 
which is for this ensemble $Z_V=0.627(4)$ \cite{Alexandrou:2012mt}. 
After renormalization the condition $F_1^{u-d}(Q^2=0)=1$ (see e.g. 
\cite{Hagler:2009ni}) is fulfilled within errors.

\section{Conclusion and outlook}

We have investigated a new method for the computation of 
quasi parton distributions and have shown that using a stochastic estimator 
for the all-to-all propagator is a well-suited method for the computation 
of the necessary matrix elements. Smearing is shown not to improve the 
statistical uncertainty and was thus dropped. 

Given the first results obtained using 1000 measurements, we are currently increasing 
statistics, which can reach up to 30 000 measurements.

At present, we are also working on the next steps of the project.
This includes the implementation of the matching factors $Z$ 
of Eq.~(\ref{EQN_matching}). The factors were already computed
to one loop in \cite{Xiong:2013bka}, however the implementation 
and numerical solution of the equation are not easily done.
In addition we are working on implementing the
finite nucleon mass correction \cite{Steffens:2012jx}.

In parallel, we are continuing to study systematic effects
that might occur in the simulation, e.g. we study the influence
of excited states on the matrix elements by varying the source-sink
separation. We also plan to include more momenta, in particular when the 
mentioned larger statistics is available.

Another challenge is the renormalization of the quantities
that are involved. One possibility is to compute separate
renormalization factors for the matrix elements with different $\Delta z$. 
We are in the process of testing different ideas for this
renormalization process, details of which will be presented elsewhere. 

Finally, it would be attractive to use the new ETMC ensemble \cite{Abdel-Rehim:2013yaa} to compute the
quark distribution directly at the physical value of the pion mass.


\begin{thebibliography}{99}

\bibitem{Alexandrou:2014yha}
 C.~Alexandrou,
 EPJ Web Conf.\  {\bf 73} (2014) 01013
 [arXiv:1404.5213 [hep-lat]].
 
\bibitem{Ji:2013dva}
  X.~Ji,
  Phys.\ Rev.\ Lett.\  {\bf 110} (2013) 262002
  [arXiv:1305.1539 [hep-ph]].

\bibitem{Lin:2014zya}
  H.~W.~Lin, J.~W.~Chen, S.~D.~Cohen and X.~Ji,
  [arXiv:1402.1462 [hep-ph]].

\bibitem{Xiong:2013bka}
  X.~Xiong, X.~Ji, J.~H.~Zhang and Y.~Zhao,
  Phys.\ Rev.\ D {\bf 90} (2014) 014051
  [arXiv:1310.7471 [hep-ph]].

\bibitem{Alexandrou:2013xon}
  C.~Alexandrou {\it et al.}  [ETM Collaboration],
  Eur.\ Phys.\ J.\ C {\bf 74} (2014) 2692
  [arXiv:1302.2608 [hep-lat]].

\bibitem{Cichy:2012vg}
  K.~Cichy, V.~Drach, E.~Garcia-Ramos, G.~Herdoiza and K.~Jansen,
  Nucl.\ Phys.\ B {\bf 869} (2013) 131
  [arXiv:1211.1605 [hep-lat]].

\bibitem{Jansen:2009xp}
  K.~Jansen and C.~Urbach,
  Comput.\ Phys.\ Commun.\  {\bf 180} (2009) 2717
  [arXiv:0905.3331 [hep-lat]].

\bibitem{Hasenfratz:2001hp}
  A.~Hasenfratz and F.~Knechtli,
  Phys.\ Rev.\ D {\bf 64} (2001) 034504
  [hep-lat/0103029].

\bibitem{Alexandrou:2013tfa}
  C.~Alexandrou, V.~Drach, K.~Hadjiyiannakou, K.~Jansen, B.~Kostrzewa and C.~Wiese,
  
  \pos{PoS(LATTICE 2013)289}
  [arXiv:1311.3174 [hep-lat]].

\bibitem{Baron:2010bv}
  R.~Baron, P.~Boucaud, J.~Carbonell, A.~Deuzeman, V.~Drach, F.~Farchioni, V.~Gimenez and G.~Herdoiza {\it et al.},
  JHEP {\bf 1006} (2010) 111
  [arXiv:1004.5284 [hep-lat]].

\bibitem{Alexandrou:2012mt}
  C.~Alexandrou, M.~Constantinou, T.~Korzec, H.~Panagopoulos and F.~Stylianou,
  Phys.\ Rev.\ D {\bf 86} (2012) 014505
  [arXiv:1201.5025 [hep-lat]].

\bibitem{Hagler:2009ni}
  P.~Hagler,
  Phys.\ Rept.\  {\bf 490} (2010) 49
  [arXiv:0912.5483 [hep-lat]].

\bibitem{Steffens:2012jx}
  F.~M.~Steffens, M.~D.~Brown, W.~Melnitchouk and S.~Sanches,
  Phys.\ Rev.\ C {\bf 86} (2012) 065208
  [arXiv:1210.4398 [hep-ph]].


\bibitem{Abdel-Rehim:2013yaa}
  A.~Abdel-Rehim, P.~Boucaud, N.~Carrasco, A.~Deuzeman, P.~Dimopoulos, R.~Frezzotti, G.~Herdoiza and K.~Jansen {\it et al.},
  \pos{PoS(LATTICE 2013)264}
  [arXiv:1311.4522 [hep-lat]].

\end{thebibliography}
\end{document}